\documentclass[conference]{IEEEtran}
\IEEEoverridecommandlockouts

\usepackage{cite}
\usepackage{amsmath,amssymb,amsfonts}
\usepackage{algorithmic}
\usepackage{subfigure}
\usepackage{graphicx}
\usepackage{textcomp}
\usepackage{xcolor}
\usepackage{gensymb}

\usepackage{amsmath}
\usepackage{siunitx}
\usepackage{booktabs,colortbl, array}
\usepackage{pgfplotstable}
\pgfplotsset{compat=1.8}
\definecolor{rulecolor}{RGB}{0,71,171}
\definecolor{tableheadcolor}{gray}{0.92}
\usepackage[font=small,labelfont=bf]{caption}

\newcommand{\topline}{ %
        \arrayrulecolor{rulecolor}\specialrule{0.1em}{\abovetopsep}{0pt}%
        \arrayrulecolor{tableheadcolor}\specialrule{\belowrulesep}{0pt}{0pt}%
        \arrayrulecolor{rulecolor}}
\newcommand{\midtopline}{ %
        \arrayrulecolor{tableheadcolor}\specialrule{\aboverulesep}{0pt}{0pt}%
        \arrayrulecolor{rulecolor}\specialrule{\lightrulewidth}{0pt}{0pt}%
        \arrayrulecolor{white}\specialrule{\belowrulesep}{0pt}{0pt}%
        \arrayrulecolor{rulecolor}}
\newcommand{\bottomline}{ %
        \arrayrulecolor{white}\specialrule{\aboverulesep}{0pt}{0pt}%
        \arrayrulecolor{rulecolor} %
        \specialrule{\heavyrulewidth}{0pt}{\belowbottomsep}}%

\newcommand{\midheader}[2]{%
        \midrule\topmidheader{#1}{#2}}
\newcommand\topmidheader[2]{\multicolumn{#1}{c}{\textsc{#2}}\\%
                \addlinespace[0.5ex]}

\pgfplotstableset{normal/.style ={%
        header=true,
        string type,
        column type=l,
        every odd row/.style={
            before row=
        },
        every head row/.style={
            before row={\topline\rowcolor{tableheadcolor}},
            after row={\midtopline}
        },
        every last row/.style={
            after row=\bottomline
        },
        col sep=&,
        row sep=\\
    }
}

\def\BibTeX{{\rm B\kern-.05em{\sc i\kern-.025em b}\kern-.08em
    T\kern-.1667em\lower.7ex\hbox{E}\kern-.125emX}}
\begin{document}

\title{Dynamic User Grouping based on Location and Heading in 5G NR Systems\\

\thanks{This work is partially sponsored by the Swedish Foundation for Strategic Research and Ericsson AB.}
}

\author{\IEEEauthorblockN{Dino Pjanić, Korkut Emre Arslantürk, Xuesong Cai, Fredrik Tufvesson}
\IEEEauthorblockA{\textit{Dept. of Electrical and Information Techology, Lund University; Ericsson AB, Lund, Sweden} \\
email: dino.pjanic@ericsson.com, kemrearslanturk@gmail.com, xuesong.cai@eit.lth.se, fredrik.tufvesson@eit.lth.se}
}

\maketitle

\begin{abstract}
User grouping based on geographic location in fifth generation (5G) New Radio (NR) systems has several applications that can significantly improve network performance, user experience, and service delivery. We demonstrate how Sounding Reference Signals channel fingerprints can be used for dynamic user grouping in a 5G NR commercial deployment based on outdoor positions and heading direction employing machine learning methods such as neural networks combined with clustering methods.

\end{abstract}

\begin{IEEEkeywords}
5G, beamforming, localization, machine learning, positioning, radio access network, Sounding Reference Signal, user grouping, 
\end{IEEEkeywords}

\section{Introduction}
\label{section:Intro}
The need for User Equipment (UE) positioning in cellular networks dates back to their early generations, initially driven by requirements for emergency call localization. Precise geographical localization capabilities have been a subject of research for decades. Although most existing localization solutions are enabled by Global Navigation Satellite Systems (GNSS), there is an increasing need for standalone positioning capabilities within fifth-generation (5G) cellular systems. GNSS technology can be unreliable in dense urban environments due to shadowing, multipath propagation, and poor satellite coverage \cite{GNSS, GNSS2}. Driven by various use cases such as smart factories, autonomous vehicles, and sensing, cellular UE positioning has emerged as a key service provided by 5G networks. Recent research has advanced 5G outdoor positioning to very accurate solutions, broadly classified into two categories: conventional signal processing \cite{Fredrik, Xuesong, Fredrik2}, and Machine Learning (ML) based methods \cite{Butt, Burghal, Russ}. Signal processing methods, which use Time of Arrival (ToA), Angle of Arrival (AoA), and Time Difference of Arrival (TDoA), require the estimation of radio channel parameters between UE and base stations (BS). In contrast, ML-based methods rely on pre-processed data for training.  \newline 
\indent However, many essential 5G functionalities, such as mobility management, network planning, and data analytics \cite{LocBasedAnalytics} cannot depend on GNSS services and must rely on internal positioning capabilities. As positioning techniques in 5G new radio (NR) systems evolve, position-based user grouping becomes the next logical step, enhancing the ability to capture spatial relationships alongside grouping based on channel conditions. This approach could significantly benefit key 5G functions, including:
\begin{itemize}
    \item \emph{Network Resource Optimization}: Load balancing can use geographical UE grouping to comprehend UE distribution across different cells, avoiding congestion. Spectrum efficiency improves by grouping UEs based on cell location, allowing more efficient frequency reuse and interference management.  
    \item \emph{Enhanced Mobility Management}: Handover optimization benefits from understanding the movement patterns and behavior of the UE. The network can anticipate handovers and prepare target cells in advance, reducing handover failures and maintaining service continuity.
    Context-aware services can provide UEs with relevant information and services based on their current geographical position.
    \item \emph{Quality of Service Improvement}: Using beamforming in targeted areas, beamforming techniques can be optimized to enhance signal quality and data rates.
     \item \emph{Network Planning}: Infrastructure deployment can leverage geographical data to guide the placement of new BSs, small cells, and other network infrastructure. Capacity planning can anticipate areas and times of high UE densities, helping to manage demand.
\end{itemize}

\indent There are significant research gaps in location-based UE clustering within the 5G NR system. The authors of \cite{Dino} explored the classification of UEs in dynamic millimeter-wave scenarios using conventional ML techniques on simulated CSI-RS measurements without directly considering physical positioning. The study in \cite{DynamicUEGroup} proposes dynamic UE-group-based interference management by adjusting data transmission powers in small cell deployments. As a reference, our study utilizes a high-resolution uplink (UL) Sounding Reference Signal (SRS) dataset, recently showcased in a highly accurate positioning model \cite{GuodaDino}. This model outperforms previous studies by regressing UE positions using an attention-based approach. The major contributions of this paper are as follows:
\begin{itemize}
    \item We propose an accurate outdoor positioning model that utilizes SRS channel estimates to infer the actual user position and the Course Over Ground (COG) or heading direction.
     \item To the best of our knowledge, this is the first study to introduce geographical-based user grouping through clustering methods in a commercial 5G system.
\end{itemize}

\section{System Model and Data Collection}
\label{System}
The fundamental concept behind positioning with UL SRS channel estimates is that a specific physical location under similar radio channel conditions corresponds to unique SRS-generated channel estimates. Each UE transmit antenna acts as a resource where channel measurements are gathered by the numerous antennas of the massive Multiple-Input Multiple-Output (MIMO) receiver in the UL. At the BS end, a MIMO system enables improved channel measurements across multiple frequency resource instances of the entire bandwidth, incorporating the UE location information via AoA, ToA and TDoA.
We consider a commercial 5G NR Time Division Duplex (TDD) system in a single-user massive MIMO scenario, where the BS processes a time series of SRS measurements that capture the angular delay spectrum of the radio channel in the beam domain. At time~$t$, the UE, equipped with $M_{\text{UE}}$ antenna elements, transmits an UL pilot signal. This pilot signal reaches the BS at an azimuth arrival angle $\phi$ and an elevation angle $\theta$. 
The BS is equipped with $M_{\text{BS}}$ antennas, half of which is vertically polarized and the other half horizontally polarized. We suppose that the number of multipath components is $P$, and denote $\tau_{p,t}$ as the time delay between UE and BS w.r.t. the $p$-th path at time $t$, and $\zeta_{p,m,t}$ denotes the complex amplitude of each multipath component. The BS utilizes all vertical-polarized antennas to form $N_{\text{V}}$ beams, the response of the $i$-th beam w.r.t. the $p$-th path is $\psi_{\text{V},i}(\phi_p, \theta_p, f)$, where $f$ denotes frequency, and $\phi_p$ and $\theta_p$ represent the azimuth and elevation arrival angles for the $p$-th multipath, respectively. Another set of $N_{\text{H}}$ beams uses all horizontal polarized antennas, and the response of the $i$-th beam is $\psi_{\text{H},i}(\phi_p, \theta_p,f)$. 
For the $m$-th UE antenna, the propagation channel is modeled for each beam at time index $t$ as 
\begin{equation}
\begin{aligned}
     h_{\text{V},i,m,t}(f) &= \sum_{p=1}^P \psi_{\text{V},i}(\phi_p, \theta_p, f)\hspace{1pt}\zeta_{p,m,t}\hspace{1pt}\exp\{-j2\pi\hspace{1pt}f\hspace{1pt}\tau_{p,t}\} \\
      h_{\text{H},i,m,t}(f) &= \sum_{p=1}^P \psi_{\text{H},i}(\phi_p, \theta_p, f)\hspace{1pt}\zeta_{p,m,t}\hspace{1pt}\exp\{-j2\pi\hspace{1pt}f\hspace{1pt}\tau_{p,t}\}.
\end{aligned}
\end{equation}
By collecting all $h_{\text{V},i,m,t}(f)$ and $ h_{\text{H},i,m,t}(f)$ for the $F$ subcarriers, we formulate two beam space matrices of the Channel Transfer Functions (CTFs), $\mathbf{H}_{\text{V},m,t} \in \mathbb{C}^{N_{V} \times F}$ and $\mathbf{H}_{\text{H},m,t} \in \mathbb{C}^{N_{\text{H}} \times F}$ at time $t$, which correspond to the vertical and horizontal polarized antenna groups, respectively. We further define the matrix $\mathbf{H}_t \in \mathbb{C}^{N \times F} = \left[\mathbf{H}_{\text{H},1,t}^T,\mathbf{H}_{\text{V},1,t}^T,..., \mathbf{H}_{\text{H},M_{\text{UE}},t}^T,\mathbf{H}_{\text{V},M_{\text{UE}},t}^T\right]^T$ that combines the channel matrices of all UE antennas as $N = M_{\text{UE}}\hspace{1pt}(N_{\text{H}}+N_{\text{V}})$ that depends on the UE position and therefore meets the criteria needed to perform ML-based localization.

\subsection{Outdoor 5G NR measurement campaign}
\begin{figure}[t]
	\centering  
    \includegraphics[width=1.0\linewidth]{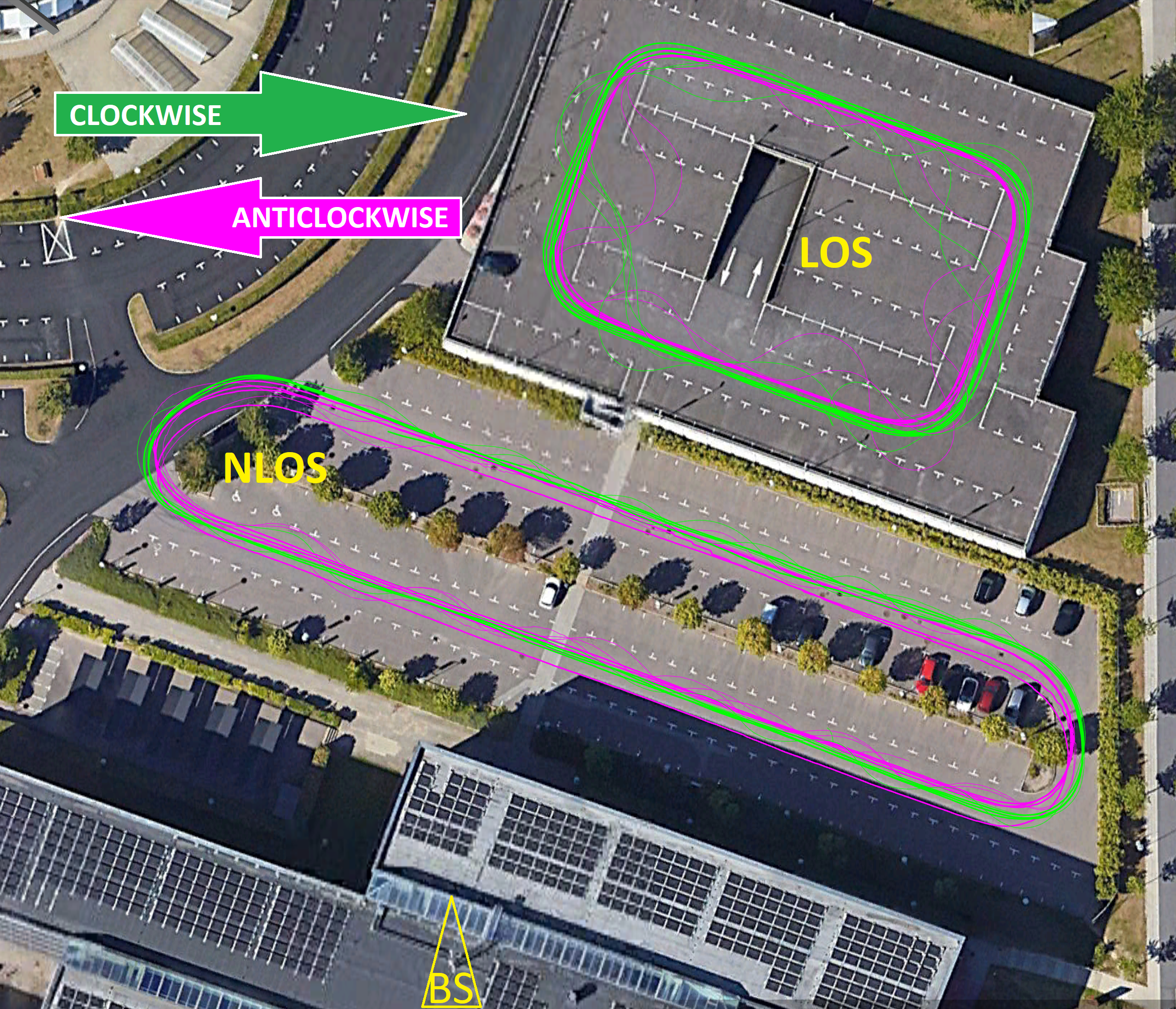}
	\caption{The 5G NR base station was installed on top of a 20-meter-high building. During this measurement campaign, a test vehicle traversed two predefined routes: A 10-meter-high garage path for LoS measurements and a ground-level path for NLoS measurements below the base station building. Each route features four different movement patterns. Thinner lines depict random trajectories.}
	\label{fig:MeasurementCampaign} 
\end{figure}

To assess our localization pipeline, we conducted an outdoor measurement campaign in a parking lot near the Ericsson office in Lund, Sweden. Fig.~\ref{fig:MeasurementCampaign} shows photos of the BS antenna and measurement locations. 
Throughout the campaign, a commercial UE was placed on top of a test vehicle alongside a high-performance GNSS receiver, providing ground truth reference with centimeter-level positioning accuracy and COG parameter featuring GNSS multi-band and multi-constellation support. To ensure uninterrupted SRS transmission, the UE remained in a connected state while simultaneously downloading data at a rate of 750 Mbit/s. The UL SRS pilot signals were received and processed by a commercial Ericsson 5G BS in TDD mode, operating in the mid-band at a center frequency of $3.85$ GHz, compliant with the 5G NR 3GPP standard 38.104 Rel15 \cite{38104}. The BS was equipped with an integrated radio with 64 transmitters/receivers (TX/RX) and $32$ dual polarized antennas. For digital beam forming, the 64 TX/RX formulate 64 beams in DL/UL respectively. As illustrated in Fig. \ref{fig:MeasurementCampaign}, our measurement campaign includes two distinct scenarios: Line-of-Sight (LoS) and non Line-of-Sight (NLoS). In both scenarios, the velocity of the test vehicle was approximately $5$ m/s. The trajectory for each of the two measurement scenarios consists of 4 laps with 4 different UE mobility patterns: clockwise, clockwise random, anticlockwise, and anticlockwise random. This approach creates four distinct movement trajectories for each scenario, which makes them suitable for clustering. As UEs move, the clockwise and anticlockwise patterns cause them to dynamically move towards or away from each other, while the random trajectories introduce additional variation to the geographical distribution of the UEs. To increase the total number of UEs in each scenario, the data generated during the 4 laps were divided in half, resulting in an additional 4 \emph{virtual} users, totaling 8 users with 2 laps in each scenario.

\subsection{Signal Processing Pipeline}
\begin{figure}[htbp]
	\centering  
    \includegraphics[width=1.15\linewidth]{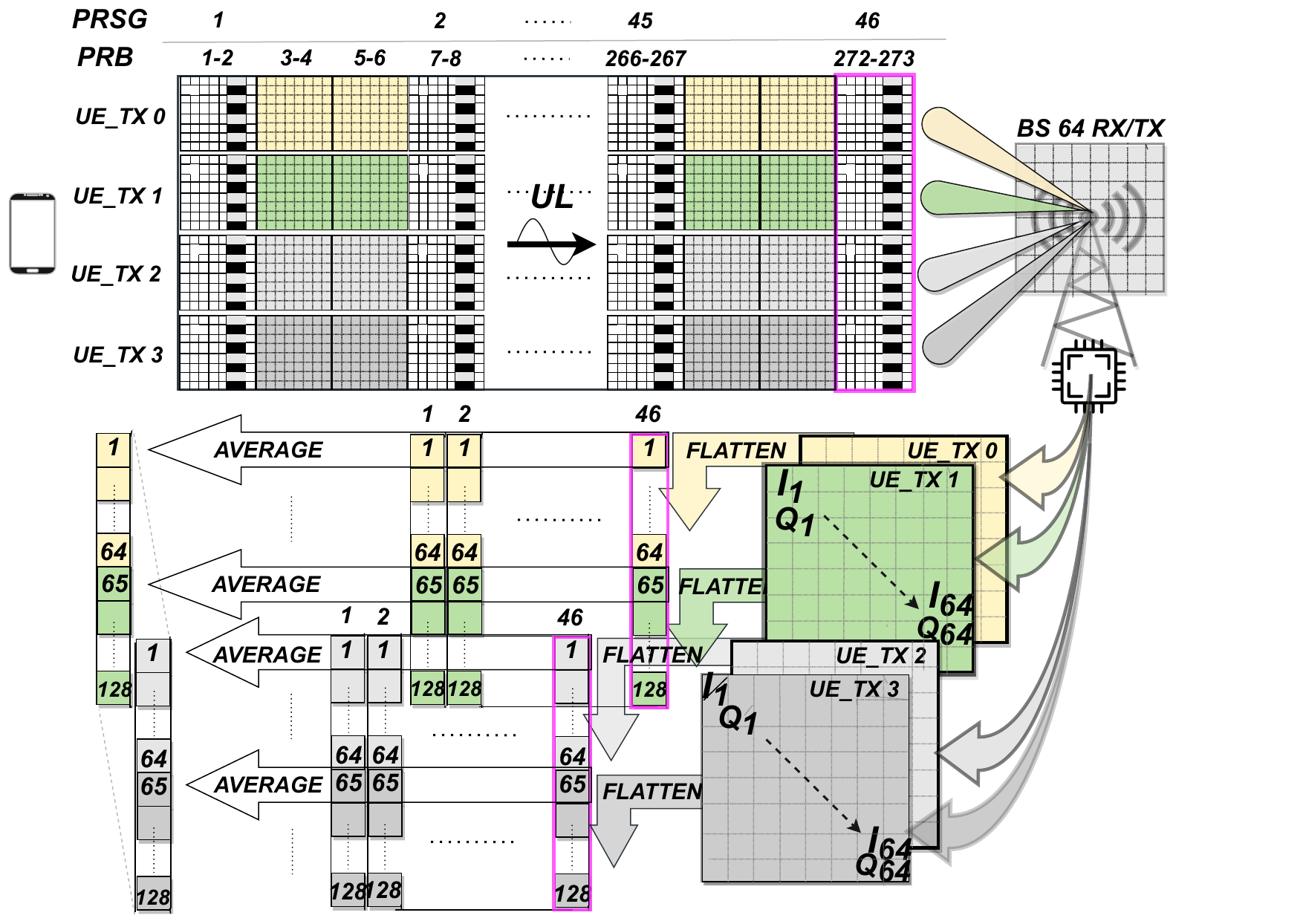}
	\caption{SRS data stream collection and pre-processed CTF dataset.}
	\label{fig:SRS-structure} 
\end{figure}

As illustrated in Fig. \ref{fig:SRS-structure}, the SRS channel estimates cover 273 Physical Resource Blocks (PRBs) over a 100 MHz bandwidth. Each channel snapshot comprises 273 PRBs for all 64 beams based on SRS reporting periodicity of 20 ms. 
The PRBs are grouped in adjacent pairs and averaged by downsampling, taking the mean value of the sampled data points resulting in 137 PRB Subgroups (PRSGs). Downsampling was performed on every third PRSG, resulting in a total of 46 PRSGs. The UE, equipped with 4 antennas (i.e., 4 UE layers), transmits the SRS pilots. We recorded the SRS pilots from 2 UE layers at a time, forming two channel transfer function matrices $\mathbf{H}_1, \mathbf{H}_2 \in \mathbb{C}^{N \times F}$. We define a matrix $\mathbf{H}' \in \mathbb{C}^{2N \times F}$ to collect those two matrices, specifically, $\mathbf{H}' = [\mathbf{H}_1, \mathbf{H}_2]$ (with $N = 64, F = 46$).  After initial processing, the final CTF snapshot for 64 gNodeB antennas and two UE layers has a dimension of 1x128 amplitude instances, as depicted in Fig. \ref{fig:SRS-structure}, collected and averaged over 46 PRSGs for each gNodeB antenna. Gathering UL SRS channel measurements in a commercial 5G NR BS faces constraints when retrieving data-rich structures such as SRS channel measurement samples. The extensive SRS data, generated at millisecond intervals, typically reside within the BS's baseband entity, primarily for internal processing. However, accessing these data externally may be impeded by hardware and software constraints. As not all PRSG values are updated during SRS transmission, it is necessary to represent the missing channel estimate values. To ensure the validity of the CTF for missing PRSGs, we employ the simplest method, such as forward-filling, using the latest known values.

\section{Proposed ML-based Clustering Framework} 
\label{MLModels}
\begin{figure}[b]
	\centering  
        \includegraphics[width=\linewidth]{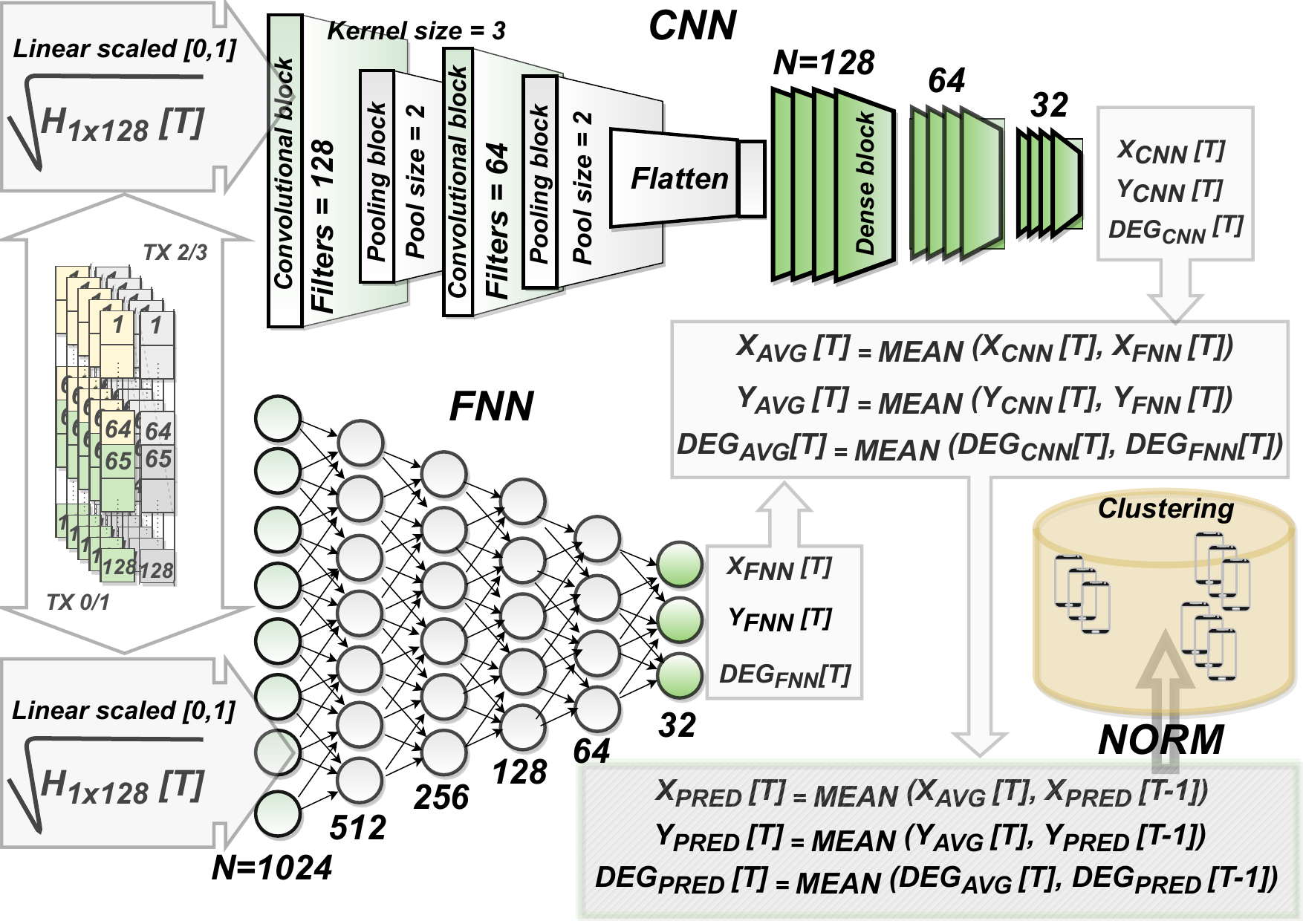}
	\caption{The architecture of the ML-driven grouping framework along with the input pipeline and intended output. The output of the positioning model is averaged with the previous prediction as a post-processing step and normalized before forwarding it to the clustering algorithms for user grouping.}
	\label{fig:ML_model} 
\end{figure}

ML-driven UE grouping framework is illustrated in Fig.  \ref{fig:ML_model} consisting of two sequential blocks:
\begin{enumerate}
\item \textbf{Positioning block:} This block is designed to achieve precise positioning and incorporates a Convolutional Neural Network (CNN) \cite{CNN} in conjunction with a Feedforward Neural Network (FNN) \cite{FNN}. Both networks utilize features extracted from the SRS dataset as input to regress the local $P_{XY}$ position, along with $COG_{DEG}$.
\item \textbf{Clustering block:} Leveraging the highly accurate positioning block, we utilize nonparametric clustering methods that do not require a pre-setting of the number of clusters, namely Density-Based Spatial Clustering of Applications with Noise (DBSCAN) \cite{DBSCAN} and Hierarchical clustering \cite{HIERARCH} for the UE grouping.
\end{enumerate}
\subsection{DBSCAN}
DBSCAN uses tree techniques called \emph{dendograms} \cite{Dendogram}, a tree-structured graph, and groups points into clusters based on their density, identifying areas with a high data point density separated by regions of low density. Since dendograms use features only indirectly as the basis for distance calculation, they partition the given data rather than entire instance space, and hence represent descriptive clustering rather than predictive one. This makes DBSCAN especially qualified for handling clusters of arbitrary shapes and sizes, even in noisy data. DBSCAN requires two parameters: epsilon (\emph{eps}) and the minimum number of samples \emph{MinPts}. The \emph{eps} represents the radius around a data point, and \emph{MinPts} the minimum number of data points within \emph{eps} to form a dense region.
\subsection{Hierarchical clustering}
Hierarchical clustering has the distinct advantage that any valid measure of distance can be used. In hierarchical clustering, deciding which clusters to combine or where to split requires a measure of dissimilarity between sets of observations. Most methods achieve this by using an appropriate distance \emph{threshold}, such as the Euclidean distance, between individual observations in the dataset. As there is a need to measure how close two clusters are, a \emph{linkage criterion} is employed, which is a general way to turn pairwise point distances into pairwise cluster distances. Our model used the WARD \emph{linkage criterion} defined as:
\begin{equation}
	   \label{eqn:Linkage}
         \Delta (A,B) = \frac{\micro_A \micro_B} {\micro_A + \micro_B} \|\vec{m_A} - \vec{m_B}\|^2
\end{equation}
where A and B are two sets of observations with a centre of cluster \emph{i} denoted as $\vec{m_i}$ and the number of points in it as $\micro_i$. Tables \ref{table:POS_block_params} and \ref{table:CLUST_block_params} summarize the hyperparameter settings of the ML model illustrated in Fig. \ref{fig:ML_model}.

\begin{table}[htbp]
        \centering
        \caption{Hyperparameters employed by the positioning block}
        \pgfplotstabletypeset[normal,
                columns/eg/.style={
                column name={$E_{\textup{g}}$ (\si{\electronvolt})},
                dec sep align
        }
        ]
        { Layers & Activation &  Batch s. & Epochs &  Optimizer  & Loss f. \\
        \topmidheader{5}{\textbf{FNN}}
        8 & ReLU & 64 & 200 & ADAM & MSE\\
        \midheader{5}{\textbf{CNN}}
        2 & ReLU & 64 & 200 & ADAM & MSE\\
        }
        \label{table:POS_block_params}
\end{table}

\begin{table}[htbp]
\centering
\caption{Parameters employed by the clustering block}
{\begin{center}
\begin{tabular}{ccccccc}
\hline\hline
Hyperparameter & DBSCAN \\
\hline
Distance Metric & Euclidian\\
eps & 0.5, 0.6\\ 
minPTS  & 1 \\ 
Algorithm  & Auto \\
\hline\hline
Hyperparameter & HIERARCHICAL \\
\hline
Distance Metric & Euclidian\\
Distance Threshold & 0.5, 1.0\\ 
Compute Full Tree & Auto \\
Linkage criterion & Ward \\
\hline\hline
\end{tabular}
\end{center}
}
\label{table:CLUST_block_params}
\end{table}

\section{Results And Discussion}
\begin{figure}[htbp]
	\centering  
    \includegraphics[width=1.1\linewidth]{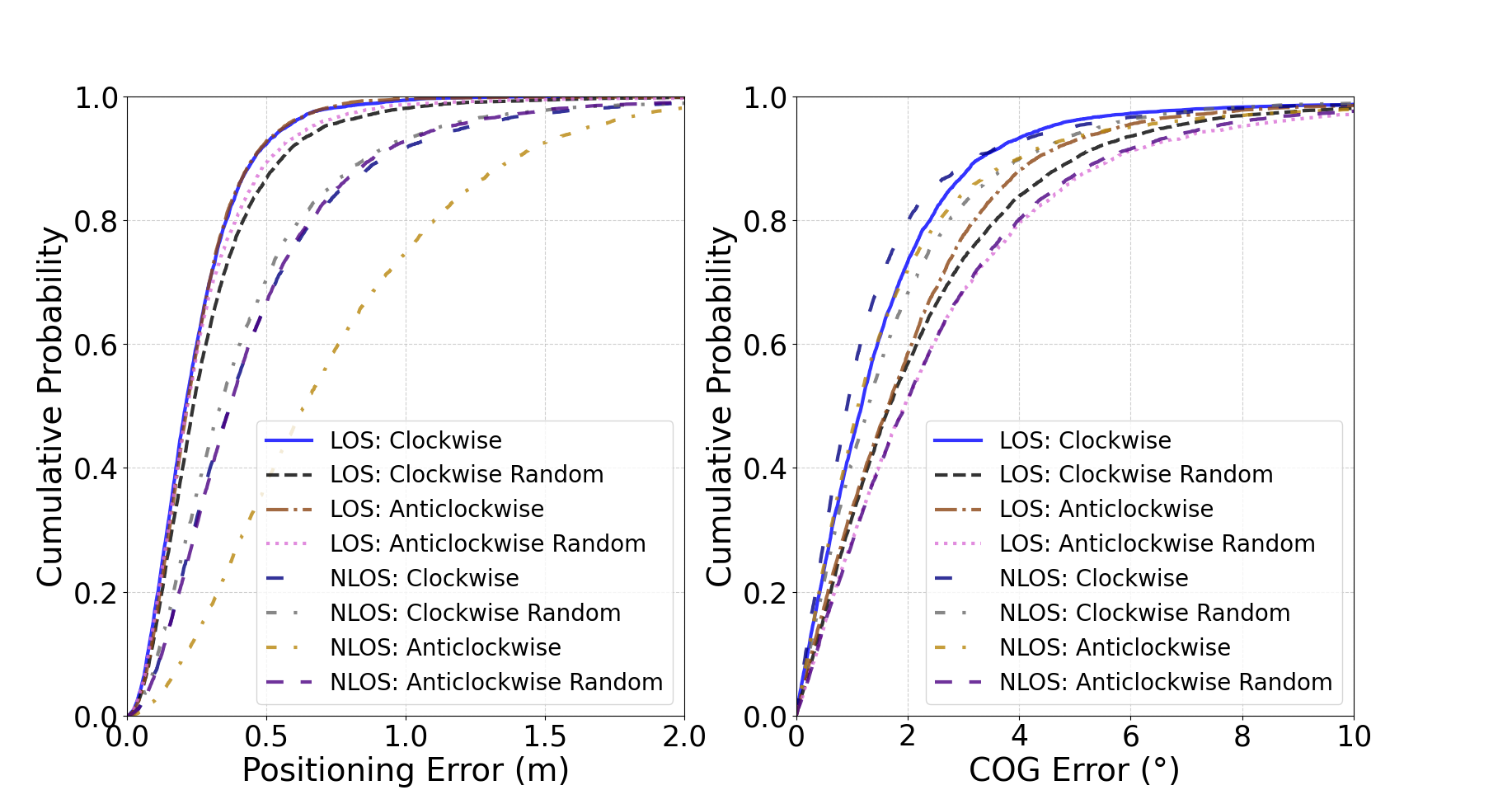}
	\caption{Positioning and COG errors.}
	\label{fig:CDF} 
\end{figure}

The accuracy of the positioning block is summarized in Table \ref{table:POS_results} and Fig. \ref{fig:CDF} demonstrating a sub-1 m accuracy level of precision and sub-9 ° heading direction.
\begin{table}[htbp]
        \centering
        \caption{Performance of the positioning block}
        \pgfplotstabletypeset[normal,
                columns/eg/.style={
                column name={$E_{\textup{g}}$ (\si{\electronvolt})},
                dec sep align
        }
        ]
        {Scenario & Metric &  X (m) & Y (m) &  Heading (°)  & Dist. (m) \\
        \topmidheader{5}{\textbf{Clockwise}}
        LoS & RMSE & 0.2  & 0.23  & 7.1 & 0.3\\
        &  R2 Score & 0.99988  & 0.99976 & 0.995 &  NA\\
        \topmidheader{5}{}
        NLoS & RMSE & 0.53 & 0.35 & 6.33 & 0.64\\
        &  R2 Score & 0.9998  & 0.9998 & 0.993 &  NA\\
        \midheader{5}{\textbf{Clockwise Random}}
        LoS & RMSE & 0.29  & 0.35 & 8.97 & 0.454\\
        &  R2 Score & 0.9997 & 0.9995 & 0.99 & NA \\
        \topmidheader{5}{}
        NLoS & RMSE &  0.52 & 0.39 & 5.8 & 0.65 \\
        &  R2 Score & 0.9997 & 0.9994 & 0.996 & NA \\
        \midheader{5}{\textbf{Anticlockwise}}
        LoS & RMSE & 0.23  & 0.21 & 8.36 & 0.31 \\
        &  R2 Score & 0.9998 & 0.9998 & 0.993 & NA \\
        \topmidheader{5}{}
         NLoS & RMSE & 0.53 & 0.75 & 8.76 & 0.92 \\
        &  R2 Score & 0.9997 & 0.998 & 0.991 & NA \\
        \midheader{5}{\textbf{Anticlockwise Random}}
         LoS & RMSE & 0.26 & 0.26 & 8.56 & 0.36 \\
        &  R2 Score & 0.9998 & 0.9997 & 0.993 & NA \\
        \topmidheader{5}{}
        NLoS & RMSE & 0.48 & 0.44 & 8.37 & 0.65 \\
        &  R2 Score & 0.9998 & 0.9991 & 0.991 & NA \\
        }
        \label{table:POS_results}
\end{table}

\begin{figure}[htbp]
\begin{minipage}[h]{0.49\linewidth}
\includegraphics[width=1\linewidth]{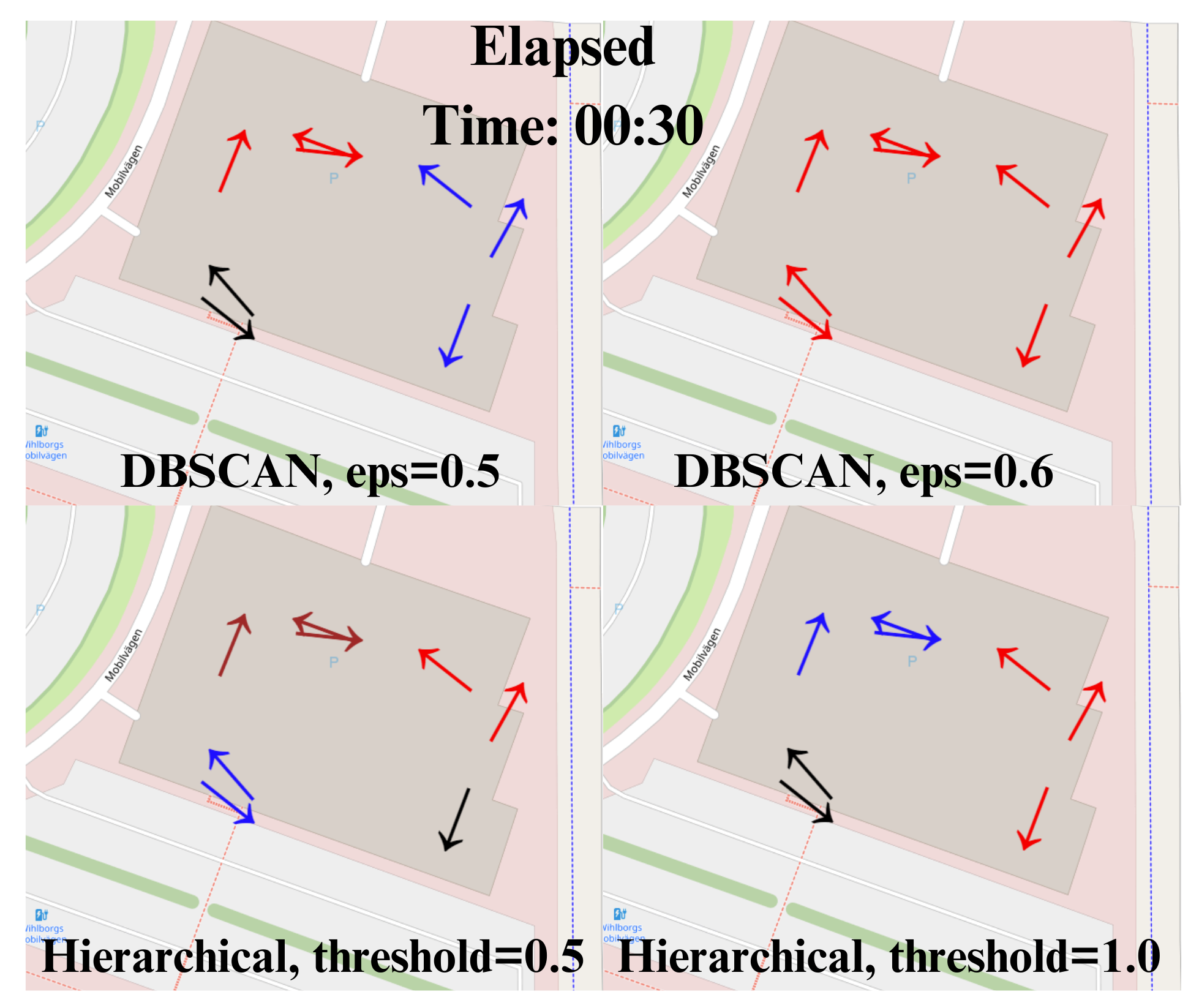}
\end{minipage}
\hfill
\begin{minipage}[h]{0.49\linewidth}
\includegraphics[width=1\linewidth]{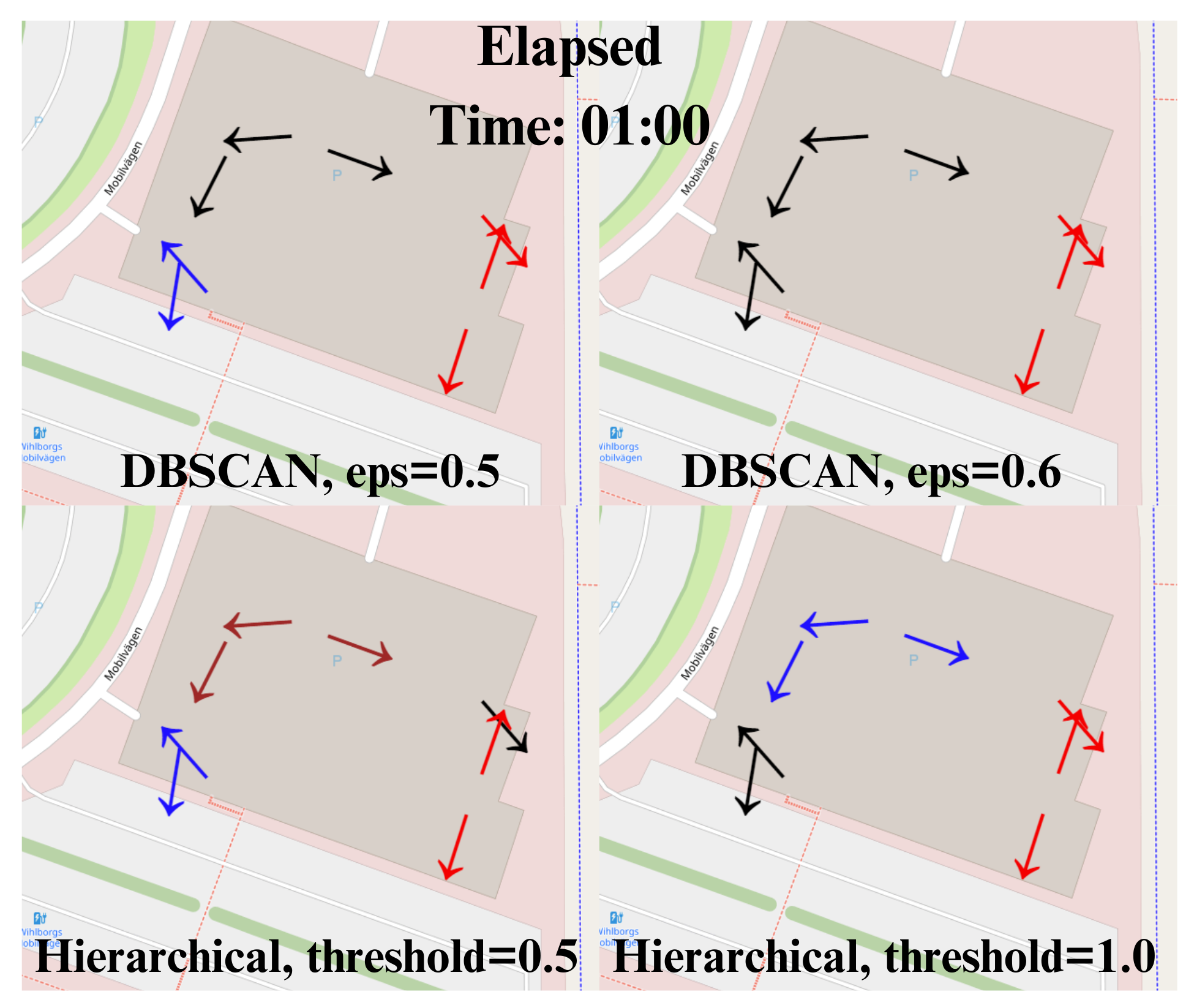}
\end{minipage}
\caption{LoS clustering results based on position}
\label{fig:LOS_clustering_pos}
\end{figure}

\begin{figure}[htbp]
\begin{minipage}[h]{0.49\linewidth}
\includegraphics[width=1\linewidth]{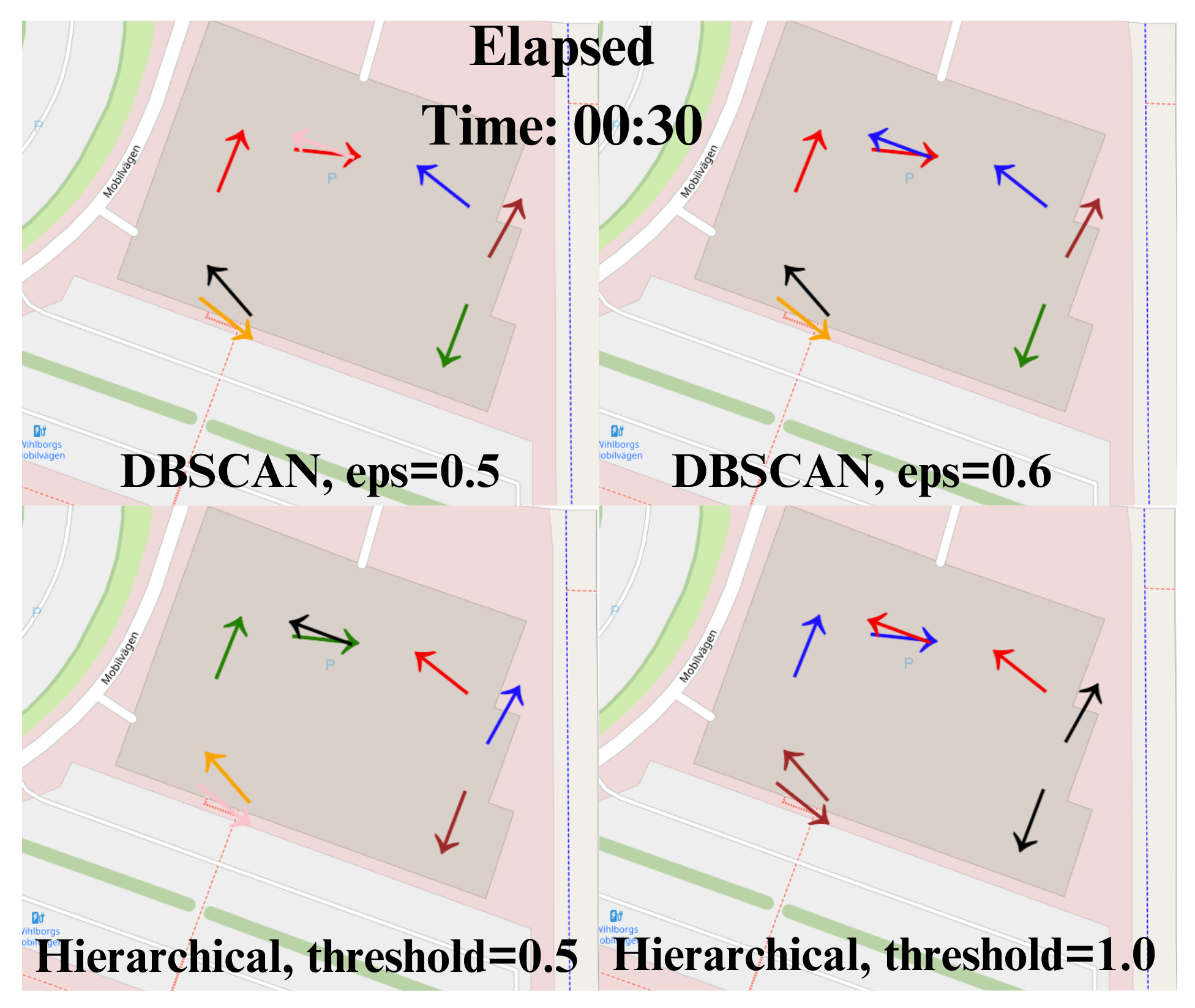}
\end{minipage}
\hfill
\begin{minipage}[h]{0.49\linewidth}
\includegraphics[width=1\linewidth]{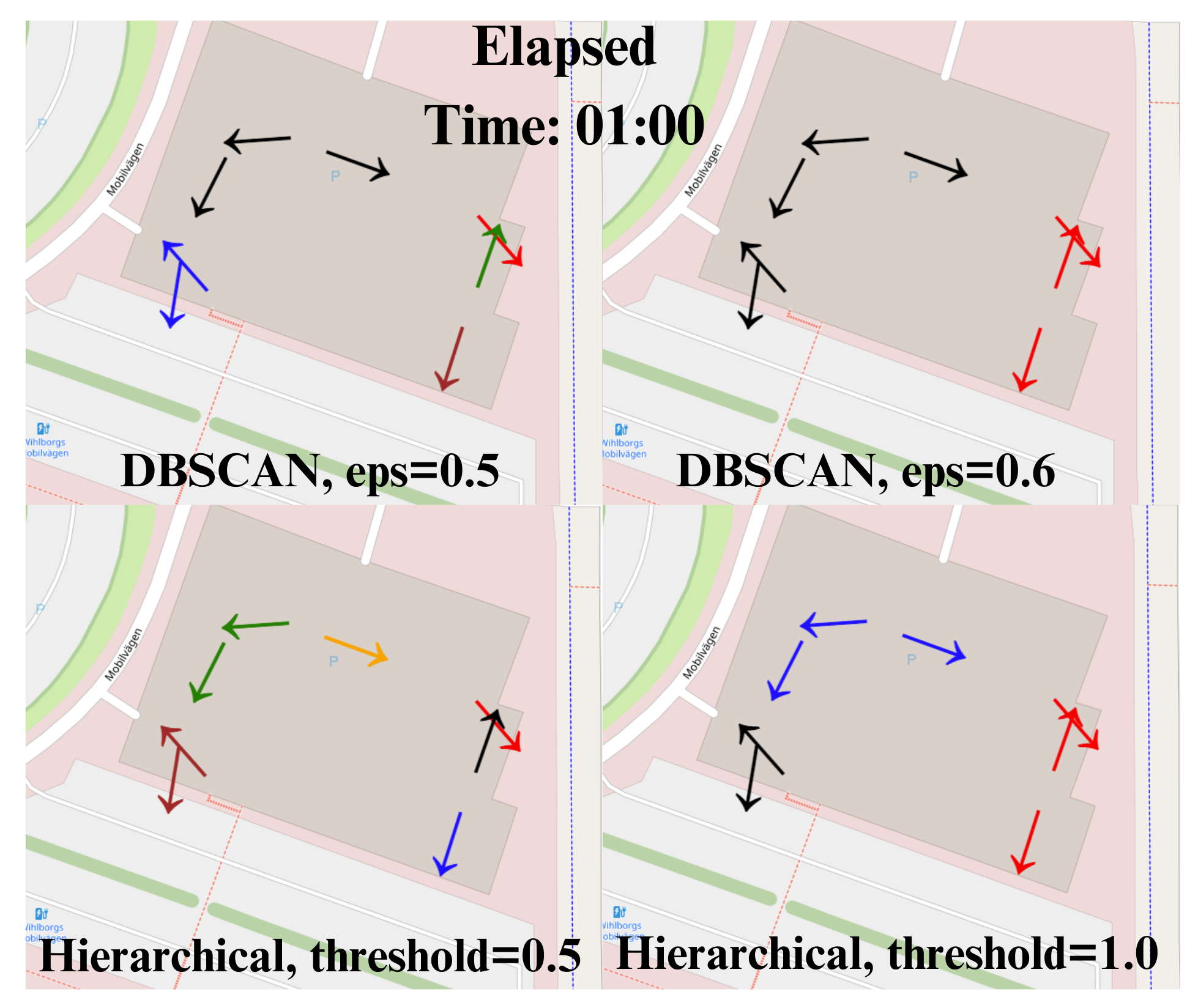}
\end{minipage}
\caption{LoS clustering results based on position and heading}
\label{fig:LOS_clustering_pos_and_head}
\end{figure}

\begin{figure}[htbp]
\begin{minipage}[h]{0.49\linewidth}
\includegraphics[width=1\linewidth]{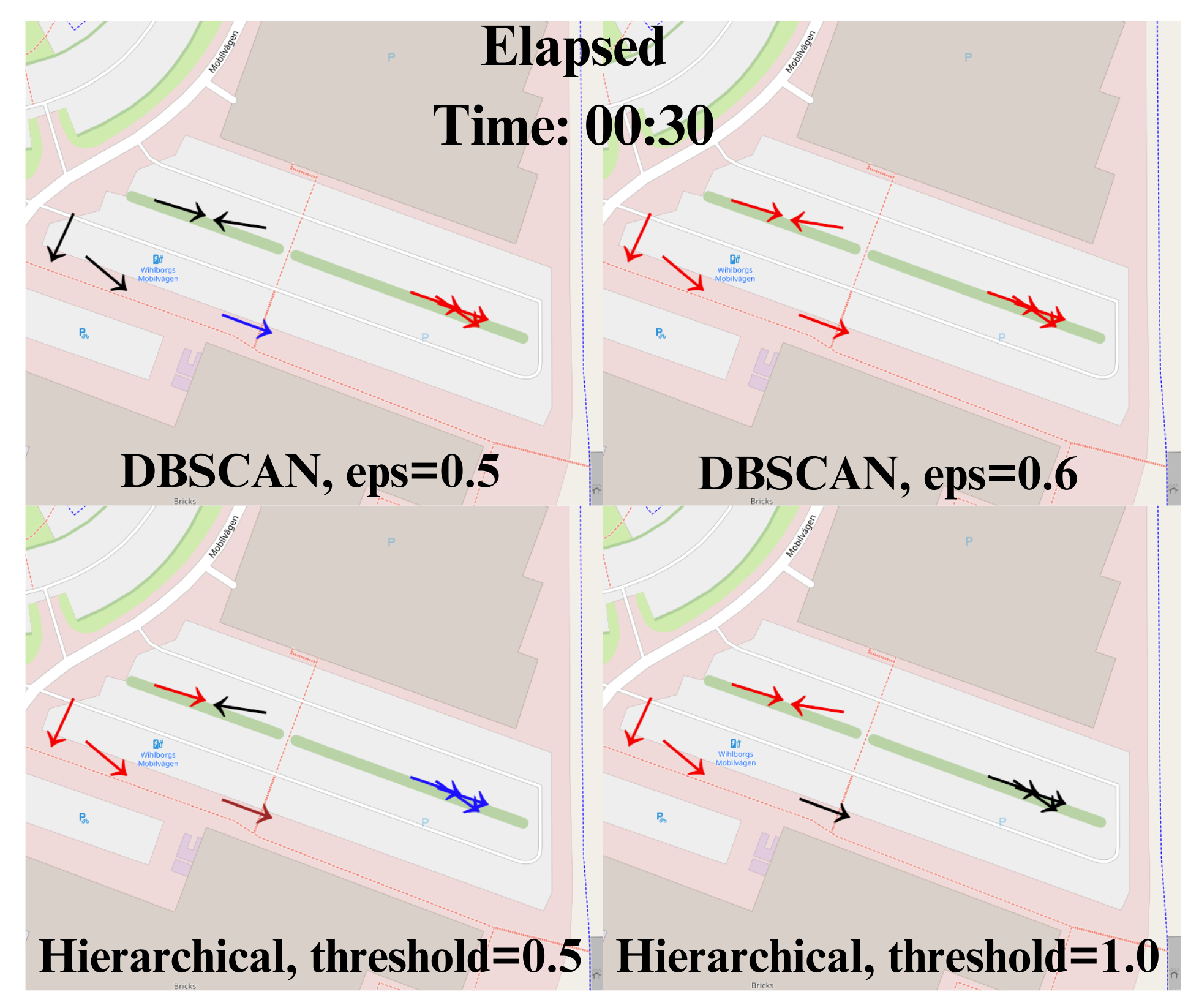}
\end{minipage}
\hfill
\begin{minipage}[h]{0.49\linewidth}
\includegraphics[width=1\linewidth]{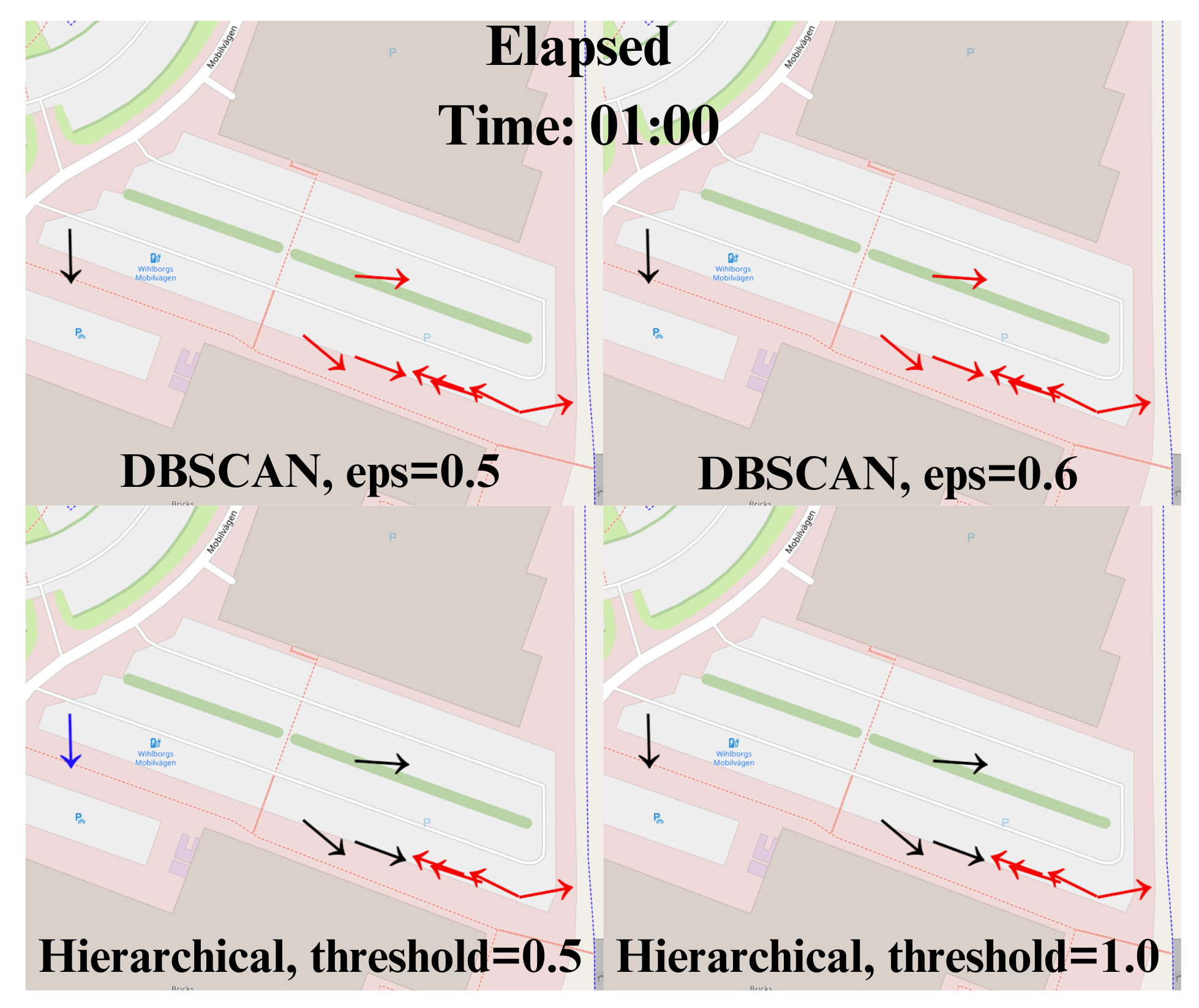}
\end{minipage}
\caption{NLoS clustering results based on position}
\label{fig:NLOS_clustering_pos}
\end{figure}

\begin{figure}[htbp]
\begin{minipage}[h]{0.49\linewidth}
\includegraphics[width=1\linewidth]{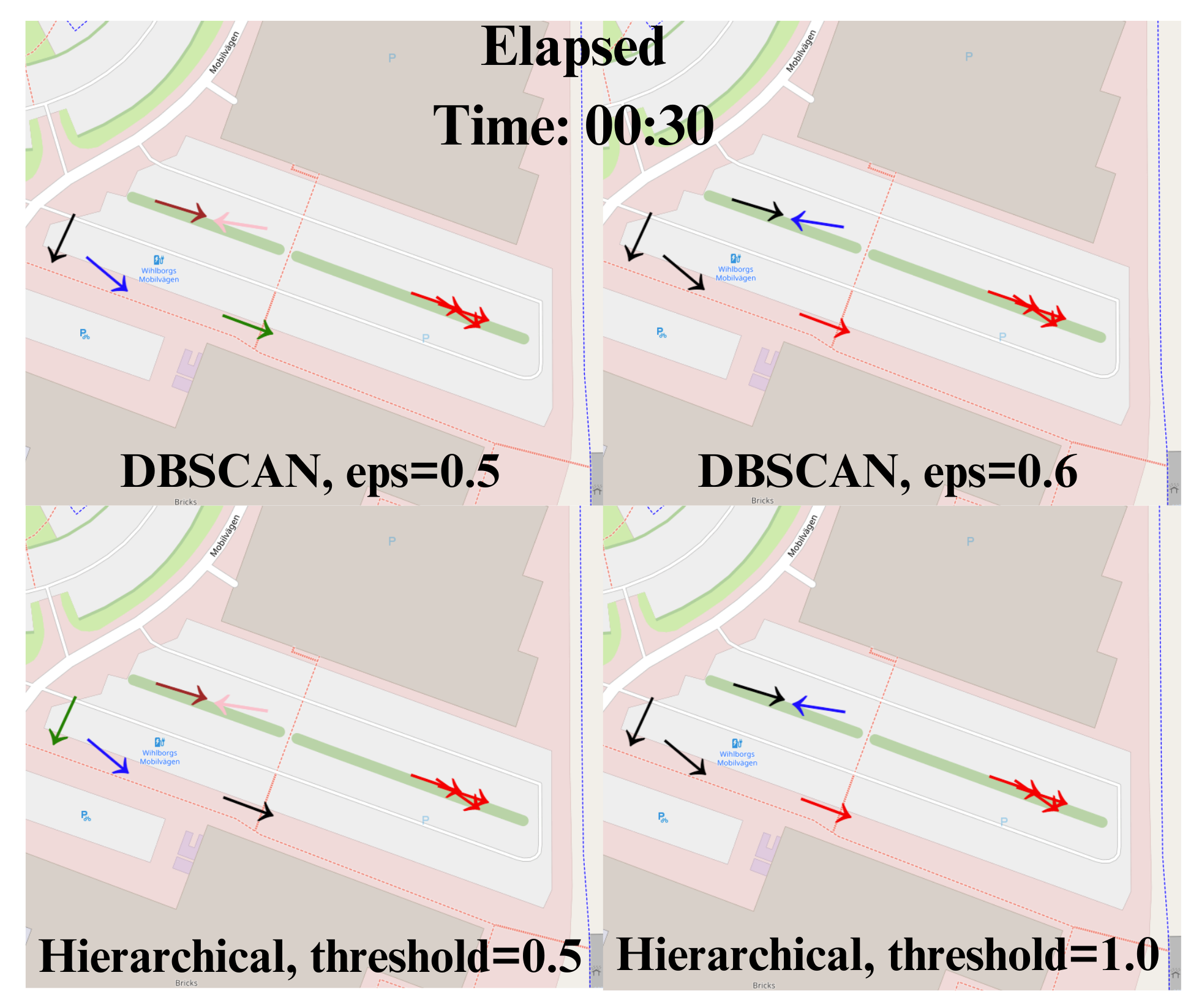}
\end{minipage}
\hfill
\begin{minipage}[h]{0.49\linewidth}
\includegraphics[width=1\linewidth]{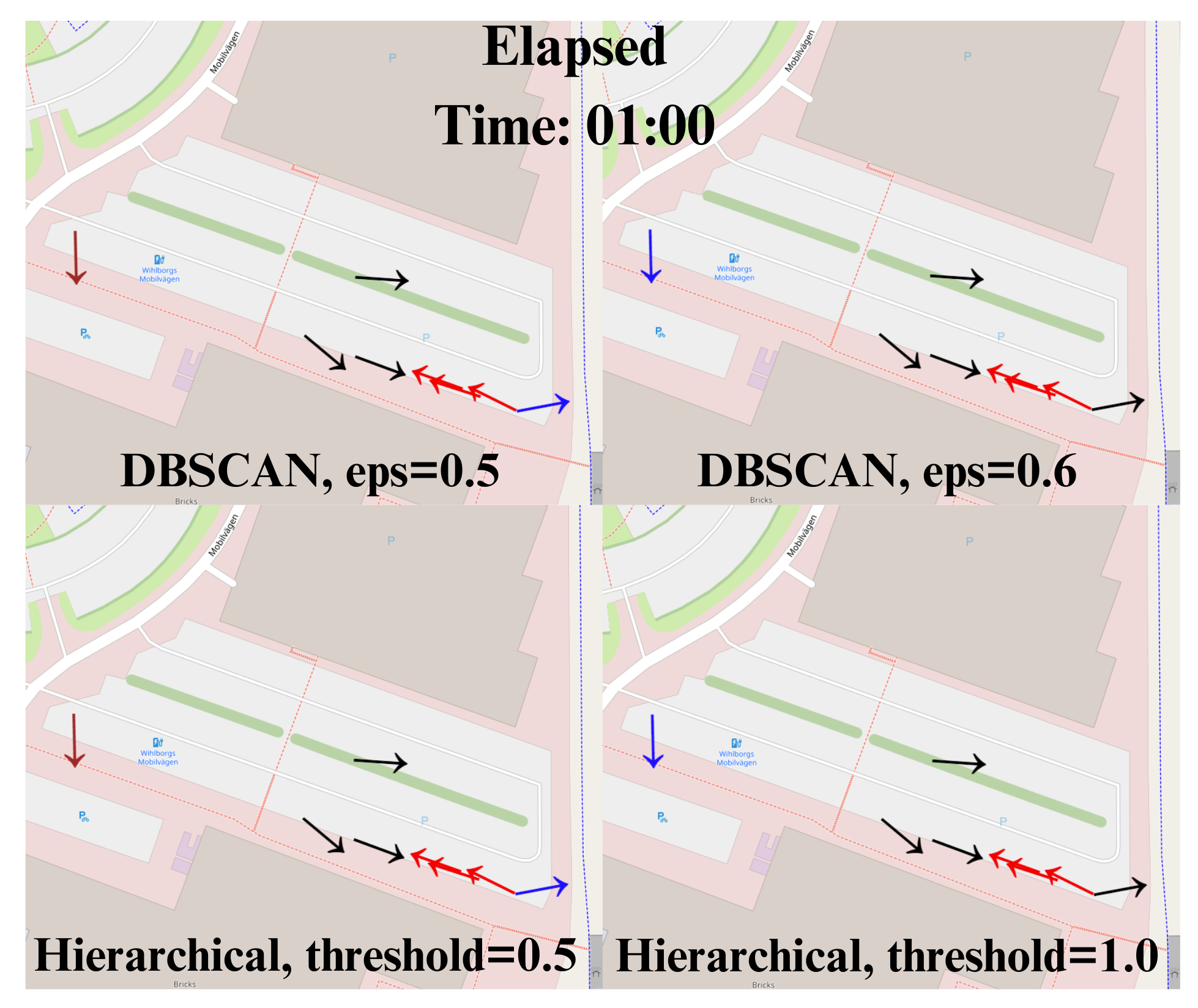}
\end{minipage}
\caption{NLoS clustering results based on position and heading}
\label{fig:NLOS_clustering_pos_and_head}
\end{figure}

\indent Based on the highly accurate performance of the proposed positioning block, the subsequent clustering block aims to partition all points in the dataset into groups of similar objects, where the notion of similarity is highly domain-dependent. As illustrated in Fig. \ref{fig:ML_model}, the two positioning features (X and Y coordinates) and the heading feature (°) were input into the clustering block. To address the dynamic nature of cellular networks, which incorporate UE mobility, non-parametric clustering algorithms using various hyperparameter values are deemed more suitable than parametric ones. Assessing the quality of user clustering in a 5G system requires understanding how well the clustering meets the specific criteria of various network functionalities. The goal of this study is not to define the best clustering hyperparameters, as these are highly dependent on specific network function domains. Therefore, we refrain from optimizing the clustering parameter settings or determining the optimality of one approach over another. Instead, we focus on demonstrating the potential of the proposed user grouping framework. We envision it as an internal capability of the 5G system, primarily for use cases requiring real-time user localization, such as beamforming, handover, or cell interference management when the UE remains connected. This also implies that users will be dynamically added to and removed from the clusters as they enter or exit the cell. \newline
\indent To underscore the heading feature's significance, we conduct user grouping based on estimated position alone and, in the subsequent step, include the heading feature. Figures \ref{fig:LOS_clustering_pos},  \ref{fig:LOS_clustering_pos_and_head}, \ref{fig:NLOS_clustering_pos}, and \ref{fig:NLOS_clustering_pos_and_head} present some time aligned comparison results of UE grouping based on various parameter settings, where each UE is represented by an arrow indicating its regressed direction and location. Colors are used to distinguish between different UE clusters. Our approach, besides the novel location-based user grouping itself, successfully incorporates the heading direction feature, which becomes particularly interesting for cases such as user movement predictions, dynamic adjustment of beamforming patterns, etc. For more detailed positioning and clustering results, we refer to the work conducted in \cite{Korkut} which used the same datasets and ML pipeline introduced in this paper.
\vspace{-22pt}
\vspace{-15pt}
\vspace{-22pt}
\vspace{-3pt}
\section{Conclusions And Future Work} 
By leveraging geographical positioning and heading direction, the proposed UE grouping framework can significantly enhance operational efficiency across various functional domains in 5G, some of which may not yet be fully realized. Future work could involve transitioning from 2-dimensional to 3-dimensional coordinate-based positioning, adding an extra dimension to the clustering method and making user grouping even more suitable for network functionalities such as beamforming. Integrating heading direction into user grouping and network management enables 5G systems to provide more intelligent location-based user grouping services.

\end{document}